\documentstyle[aps,preprint,prc]{revtex}
\makeatletter

\long\def\@caption#1[#2]#3{
  \begingroup
    \@parboxrestore
    \small
    \baselineskip=12pt
    \advance\leftskip by 1cm
    \advance\rightskip by 1cm
    \@makecaption{\csname fnum@#1\endcsname}{\ignorespaces
     {#3}}\par
  \endgroup}
\makeatother
\begin{document}
\baselineskip=24pt

\begin{center}
{\Large \bf Parity Mixed Doublets in A = 36 Nuclei}\\

\vspace{ 24pt}

{\bf Mihai Horoi}

\vspace{ 20pt}
{\sl National Superconducting Cyclotron Laboratory
and\\
Department of Physics and Astronomy,
Michigan State University,\\
East Lansing, Michigan 48824-1321, USA\\}

\vspace{.3cm}
and
\vspace{.3cm}

{\sl Department of Theoretical Physics, Institute of Atomic Physics,\\
 Magurele, Bucharest, POBox MG - 6, R - 76900, Romania\\}

\end{center}

\vspace{20pt}

\begin{abstract}

The $\gamma$-circular polarizations ($P_{\gamma}$) and
asymmetries ($A_{\gamma}$)
of the parity forbidden M1 + E2 $\gamma$-decays:
$^{36}Cl^{\ast} (J^{\pi} = 2^{-}; T = 1; E_{x} = 1.95 $ MeV)
$\rightarrow$
$^{36}Cl (J^{\pi} = 2^{+}; T = 1; g.s.)$  and
$^{36}Ar^{\ast} (J^{\pi} = 2^{-}; T = 0; E_{x} = 4.97 $ MeV)
$\rightarrow$
$^{36}Ar^{\ast} (J^{\pi} = 2^{+}; T = 0; E_{x} = 1.97 $ MeV) are
investigated theoretically.
We use
the recently proposed Warburton-Becker-Brown shell-model
interaction.
For the weak forces we discuss comparatively different weak interaction models
based on different assumptions
for evaluating the weak meson-hadron coupling constants.
The results determine a range of $P_{\gamma}$ values from which
we find the most probable values: $P_{\gamma}$ = $1.1 \cdot 10^{-4}$ for
$^{36}Cl$ and $P_{\gamma}$ = $3.5 \cdot 10^{-4}$ for $^{36}Ar$.

\end{abstract}

\vspace{ 30pt}

\centerline {{\bf PACS numbers:} 21.60.Cs, 24.80.-x, 27.30.$+$t,
12.15.Ji}

\vspace{  0pt}

\newpage

Parity nonconservation (PNC) in the nucleon-nucleon interaction
has been observed
in the left-right asymmetry in $\vec{p}-p$ scattering \cite{KISTRYN},
in nucleon-nucleus scattering induced by polarized projectiles (such
as $\vec{p}$ \cite{ZEPS} or $\vec{n}$ \cite{FRANKLE}) in
spontaneous
$\alpha$-decay \cite{MAINZ}, \cite{CDSB}
and in the circular polarization \cite{LOBASHEV}
\cite{DGKZ}, \cite{GAR} or asymmetry
\cite{SEATTLE}, \cite{ZUERICH}, \cite{AH}, \cite{DDH}
(from polarized nuclei) of the radiation emitted in nuclear
$\gamma$-decay.
There are also
theoretical predictions for new PNC experiments in induced
$\alpha$-decay \cite{OVIDIU}, \cite{KHDC} and asymmetry of the radiation
 emitted
in nuclear $\gamma$-decay \cite{ODGS}, \cite{ODGC}.
The theoretical and experimental
work in this field has been reviewed recently
\cite{AH},
\cite{DDH},
\cite{BIZZETI}.

The controversy
\cite{AH},
\cite{DDH},
\cite{DZ},
\cite{KM},
\cite{KHA}
in calculating weak meson-nucleon coupling constants in nuclei
has greatly stimulated the investigation of possible experiments sensitive
to different components of the PNC interaction Hamiltonian (H$_{PNC}$),
that depend linearly on seven such weak coupling constants:
h$^{ \Delta T}_{meson}$: h$^{1}_{\pi}$, h$^{0}_{\rho}$, h$^{1}_{\rho}$,
h$^{2}_{\rho}$, h$^{1}_{\rho^{'}}$, h$^{0}_{\omega}$,
h$^{1}_{\omega}$.
Various linear combinations of these constants can, in principle, be extracted
in different experiments, and among these are
those for the parity mixed doublets (PMD)
\cite{AH}, \cite{ODGC}.
The most interesting PMD cases are those for which the PNC effect is enhanced
due to a small energy difference between the two states and due
to a favourable ratio of the transition probabilities.
Since the PMD has definite isospins, the transition "filters out"
specific isospin components of PNC weak interaction.

In the present paper a theoretical
investigation of two new PMD cases in nuclei with A=36
is presented. The first one, in $^{36}$Cl, is given by the
$J^{\pi}T=2^{-}1$, E$_{x}$=1.951 MeV and $J^{\pi}T=2^{+}1$, E$_{x}$=1.959 MeV
levels (see Fig. 1). The second one, in $^{36}$Ar, is given by the
$J^{\pi}T=2^{-}0$, E$_{x}$=4.974 MeV and $J^{\pi}T=2^{+}0$, E$_{x}$=4.951 MeV
levels (see Fig. 2). In order to have an amplification of the PNC effect
we look to the suppressed transitions from the $J^{\pi}T=2^{-}1$ levels to
the $J^{\pi}T=2^{+}1$ ground state for $^{36}$Cl and from $J^{\pi}T=2^{-}0$
level to the $J^{\pi}T=2^{+}0$, E$_{x}$=1.97 MeV, level for $^{36}$Ar.
The corresponding PNC matrix elements were calculated
with the shell-model code OXBASH, with
the Warburton-Becker-Brown interaction \cite{WBB}
for $2s1d$-$2p1f$ model space.
One of the cases considered here ($^{36}$Cl)
has been investigated previously \cite{ODGS}
with a much smaller
valence model space which included only the
$1d_{\frac{3}{2}}$ and $1f_{\frac{7}{2}}$ orbitals.
Within this ($1d_{\frac{3}{2}}$, $1f_{\frac{7}{2}}$)
small model space the one-body contribution to the PNC matrix element
between the members of the doublet
($M_{PNC}$) vanishes.
Within the present model
space the contribution of the one-body term  dominates the theoretical
$M_{PNC}$.
The goal of the present work is to calculate
the PNC $\gamma$ asymmetries and circular
polarizations of the proposed
gamma ray transitions within different weak-interaction models in order
judge the experimental feasibility.

The degree of circular polarization of the emitted
$\gamma$-rays is given (see Ref. \cite{BLS} chapter 9, $\S$
3 eq. (9.38)) by a sum of parity
nonconserving (PNC) and parity conserving (PC) contributions:

\begin{equation}
P_{\gamma} ( \cos \theta ) \equiv \frac{W_{right}(\theta) -
W_{left}(\theta)}{W_{right}(\theta) + W_{left}(\theta)}
= ( P_{\gamma} )_{0} \cdot R^{PNC}_{\gamma} ( \cos \theta )
+ R^{PC}_{\gamma} ( \cos \theta ),
\end{equation}

\noindent
where $R^{PC}_{\gamma}$ is a parity conserving quantity discussed
bellow,

\begin{equation}
( P_{\gamma} )_{0} =
2 \cdot \frac{M_{PNC}}{\Delta E} \sqrt{\frac{b_{+} \cdot
\tau_{-}}{b_{-} \cdot \tau_{+}} \cdot
\left( \frac{E^{-}_{\gamma}}{E^{+}_{\gamma}}\right)
^{3}}
\cdot
\sqrt{\frac{1 + \delta^{2}_{-}}{1 + \delta^{2}_{+}}}
\end{equation}

\noindent
and

\begin{eqnarray}
R^{PNC}_{\gamma} ( \cos \theta ) =
\sqrt{\frac{1 + \delta^{2}_{-}}{1 + \delta^{2}_{+}}}
\{ \sum_{\nu = 0,2,4} P_{\nu} ( cos \theta ) B_{\nu} (2)
[ F_{\nu} (1122) +
F_{\nu} (2222) \delta_{+} \delta_{+} + \nonumber \\
F_{\nu} (1222) ( \delta_{-} + \delta_{+}) ] \} \cdot
\{ \sum_{\nu = 0,2,4}
P_{\nu} ( cos \theta ) B_{\nu} (2)
[ F_{\nu} (1122) +
F_{\nu} (2222) \delta^{2}_{-} +
2 F_{\nu} (1222)  \delta_{-}]\}^{-1} .
\end{eqnarray}

\noindent
$R^{PNC}_{\gamma}$
is a multiplier due to the existence of the orientation of the
nuclear spin in the initial excited state when the mixing ratios
do not vanish.
In the above equations,
$\delta_{-}$ is the M2/E1 mixing ratio,
$\delta_{+}$ is the E2/M1 mixing ratio,
the $F_{\nu}$ coefficients are defined by

\begin{eqnarray}
F_{\nu} ( L L^{'} I^{'} I ) = (-1)^{I^{'} + 3I - 1} [(2I + 1) (2L + 1)
(2L^{'} + 1)]^{\frac{1}{2}} \nonumber \\
C ( L L^{'} \nu ;1 -1 0 ) W ( L L^{'} I I; \nu I^{'}),
\end{eqnarray}

\noindent
$C$ is the Clebsch-Gordan coefficient $C( J_1 J_2 J_3 ;M_1 M_2 M_3)$
and $W$ is the Racah coefficient.
The parity conserving (PC) quantity
is given by \cite{BLS}:

\begin{eqnarray}
R^{PC}_{\gamma}
( \cos \theta ) =
\{ \sum_{\nu = 1,3} P_{\nu} ( cos \theta ) B_{\nu} (2)
[ F_{\nu} (1122) +
F_{\nu} (2222) \delta^{2}_{-} + \nonumber \\
2 \cdot F_{\nu} (1222)  \delta_{-}]\} \cdot
\{ \sum_{\nu = 0,2,4}
P_{\nu} ( cos \theta ) B_{\nu} (2)
[ F_{\nu} (1122) +
F_{\nu} (2222) \delta^{2}_{-} +
2 F_{\nu} (1222)  \delta_{-}]\}^{-1} ,
\end{eqnarray}

\noindent
where

\begin{equation}
B_{\nu} ( 2 ) = \sum_{M} ( 2 \nu + 1 )^{\frac{1}{2}}
C(2 \nu 2; M 0 M) p(M).
\end{equation}

\noindent
$p(M)$ is the polarization fraction of the $M$-state, which
determines the degree of the orientation of the nucleus.

In order to measure a PNC effect one must find situations for
which the $R_{\gamma}^{PC}$ part in Eq. (1) vanishes. Two particular cases
have this property: i.) The case of an initially unpolarized
nucleus for which $B_{0} (2)$ = 1, $B_{\nu \neq 0} (2)$ = 0 and
$F_{0} (LL^{'} 2 2) = \delta_{LL^{'}}$.
In this particularly simple case
$P_{\gamma}$ reduces to the well known expression of the circular
polarization,  $(P_{\gamma})_{0}$.
ii.)~One may prepare a polarized state by choosing $p(M)$
= $\delta_{M 0}$ for which,
$B_{\nu = 1,3} (2) = 0$ and $R_{\gamma}^{PC}$ part vanishes.

Another observable which measures a PNC effect  is the
forward-backward asymmetry of the gamma rays emitted by polarized
nuclei

\begin{equation}
A_{\gamma}(\theta) \equiv \frac{W(\theta) - W(\pi - \theta)}
{ W(\theta) + W(\pi - \theta)}.
\label{eq:fbad}
\end{equation}

\noindent
This observable has been successfully used in the $^{19}$F case \cite{SEATTLE},
\cite{ZUERICH} in order to avoid the small efficiency of the
Compton polarimeters when one measures the degree of circular
polarization. If the mixing ratios are small
($\delta_{+},\delta_{-} \ll 1$) one can show that \cite{HC92}

\begin{equation}
A_{\gamma}(\theta) \simeq (P_{\gamma})_{0} \cdot
R_{\gamma}^{PC}(cos\theta)\ ,
\label{eq:fbaa}
\end{equation}

\noindent
where $\theta$ represents the angle between
the emitted photon and the axis of polarization (if any).
The angular distribution described by this formula has a maximum
for $\theta=0^{\circ}$ \cite{HC92}. It has the advantage that the
parity conserving (PC) circular polarization,
$R_{\gamma}^{PC}(\theta)$ in Eq. (\ref{eq:fbaa}),
can be measured experimentally.
$(P_{\gamma})_{0}$ is the essential quantity for all PNC
observable.

In order to determine the magnitude of
(P$_{\gamma})_{0}$
we have made a shell-model estimate of the PNC matrix element

\begin{equation}
M_{PNC} = < J^{\pi} T, E_{x} (MeV) \mid H_{PNC}
\mid J^{- \pi} T^{'} E^{'}_{x} (MeV)>,
\end{equation}

\noindent
where $H_{PNC}$ is the PNC Hamiltonian given
 by Desplanques, Donoghue and Holstein (DDH)
\cite{DDH},
Dubovik and Zenkin (DZ) \cite{DZ}, Adelberger and Haxton (AH) \cite{AH}
or Kaiser and Meissner (KM) \cite{KM}.

The  calculations were carried out with the shell-model code
OXBASH \cite{OXB} in the $2s1d$-$2p1f$ model space in which the
$2s_{1/2},1d_{5/2},1d_{3/2},2p_{1/2},2p_{3/2},1f_{7/2}$ and
$1f_{5/2}$
orbitals are active. The truncations we made within this model space
were
$(2s1d)^{20}$
($0\hbar \omega$)
for the positive
parity states and
$(2s1d)^{19}(2p1f)^{1}$
($1\hbar \omega$)
for the
negative parity
states.
These truncations are necessary due to the dimension
limitations, but we believe that they are realistic. The
Brown-Wildenthal interaction \cite{BRWILD}
was used for the positive parity states
and the Warburton-Becker-Brown
 interaction \cite{WBB} was used for the negative parity states.
Both interactions have been tested extensively with regards to their
reproduction of spectroscopic properties \cite{WBB}, \cite{BRWILD}.
The calculation of the
PNC matrix element which included both the core (inactive) and active
orbitals has been performed as described in Ref. \cite{BRG}.

All the components \cite{DDH}, \cite{AH} of the parity nonconserving
potential are short range
two-body operators.  Because the
behavior of the shell-model wave functions at small NN
distances has to be modified, short range correlations (SRC)
were included by multiplying
the harmonic oscillator wave functions (with $\hbar \omega$ =
($45 \cdot A^{- \frac{1}{3}}$ MeV
- $25 \cdot A^{- \frac{2}{3}}$ MeV)
by the Miller and Spencer factor \cite{MS}. This procedure is consistent
with results obtained by using more elaborate treatments of SRC
such as the generalized Bethe-Goldstone
approach \cite{DGKZ}, \cite{GAR}. The PNC pion exchange matrix is decreased
by
$30\ to\ 50 \%$ as compared with the values of the matrix elements
without including SRC, while the $\rho$ ($\omega$) exchange matrix
elements is much smaller (by a factor of $\frac{1}{3}$ to $\frac{1}{6}$).

The calculated excitation energies of the first three
T=0, $2^{+}$ levels in $^{36}$Ar are 1.927, 4.410 and 7.174 MeV. The first two
of these are in good
agreement with experimental levels at 1.970 and 4.440 MeV.
The third experimental $2^{+}$ state
at E$_{x}$=4.951 MeV (the state
belonging to the parity doublet) apparently is an intruder
state in the $2s1d$
($0\hbar \omega$)
model space.
This conclusion is also supported
by the suppressed Gamow-Teller $\beta$ transition probability
to this third state \cite{BW85}. We have thus expanded the
model space to include
some
$2\hbar \omega$
configurations - those of the type
($1d_{5/2})^{12}
(2s_{1/2},1d_{3/2})^{6}
(2p_{3/2},1f_{7/2},2p_{1/2})^{2}$.
The
$2\hbar \omega$
configurations were shifted
down by 11.5 MeV relative to the
$0\hbar \omega$ configurations
so that the first $2^{+}0$ state with a dominant
2$\hbar \omega$ component ($\sim$ 80\%) becomes the third $2^{+}$
in the calculated
$(0+2)\hbar \omega$
spectrum. The dominant PNC transition is
$1d_{3/2}$ - $2p_{3/2}$ and the DDH PNC matrix elements is 0.12
eV (see Table 1). Due to the truncations made, the PNC result
for $^{36}$Ar may not be as reliable as that for
$^{36}$Cl.

The calculated excitation energies of the first three
T=1, $2^{+}$ levels in
$^{36}$Cl
 are 0, 2.008, 2.451 and 4.429 MeV.
The first three
of these are in good
agreement with experimental levels at 0, 1.959 and 2.492 MeV.
The theoretical
B(E2) and
B(M1)
and mixing ratios are in relatively good
agreement with the experiment (see Table 1). For both
$^{36}$Ar
and $^{36}$Cl the 2$^-$ states are the lowest observed experimentally
and the theoretical wave functions should thus be fairly reliable.
The extremely weak E1 transitions
do not serve as
a useful test of the wave functions
(the one in $^{36}$Ar is isospin
forbidden).

DDH \cite{DDH} investigated a variety of
approximations within the quark model for
the weak coupling constants and discussed the model uncertainties. These
uncertainties give rise to a range of values for the PNC coupling
constants. Recently several other calculations have been made, one
within the framework of the chiral soliton model \cite{KM} and others within
the quark framework \cite{DZ}, \cite{KHA}.
Even though both of these approaches lead to
fixed values for PNC coupling constants \cite{KM} (see Table 2), the
values are subject to uncertainties. In particular, the soliton model
gives an extremely small value for $h_{\pi}$ as compared to DDH, however, this
result comes essentially from the factorization approximation, and DDH
discuss the importance of going beyond the factorization approximation
\cite{DDH}. In any case it is clear that the observation of PNC in nuclei is a
test not only of the $\Delta$-S=0 PNC component of the weak interaction,
but also of the hadronic strong interaction models.

The results (up to a complex phase factor) can be summarized as:

\begin{eqnarray}
M_{PNC} ( ^{36}Cl ) = ( 1.09 \cdot h^{1}_{\pi} -
0.20 \cdot h^{1}_{\rho} -
0.30 \cdot h^{1}_{\omega} - \nonumber \\
0.027 \cdot h^{1}_{\rho^{'}} +
0.57 \cdot h^{0}_{\rho} +
0.32 \cdot h^{0}_{\omega} +
0.015 \cdot h^{2}_{\rho} ) \cdot 10^{-2} eV,
\end{eqnarray}

\noindent
and

\begin{equation}
M_{PNC} ( ^{36}Ar ) = - ( 1.00\cdot h^{0}_{\rho} +
0.44 \cdot h^{0}_{\omega} ) \cdot 10^{-2} eV.
\end{equation}

\noindent
Here $h^{\Delta T}_{meson}$ should be given in units of 10$^{-7}$
as in Table 2.

In the $^{36}$Cl case the components containing
$h_{\pi}$
($M_{\pi}$)
and $h_{\rho ( \omega)}$
($M_{\rho + \omega }$)
couplings come in with opposite signs,
however the difference is remarkably
almost the same in all the weak interaction
models. The specific numbers are:
$M_{\pi}$ = 0.050 eV and
$M_{\rho + \omega }$  = -0.069 eV for DDH, and
$M_{\pi}$ = 0.00207 eV and
$M_{\rho + \omega }$ = -0.023 eV for KM.
In the $^{36}$Ar  case the $h_{\rho}$-components strongly dominate
the PNC matrix element ($M_{PNC}$) (e.g. within DDH: $M_{\rho}$ =
0.113 eV, while $M_{\omega}$ = 0.009 eV).

The  PNC matrix elements obtained are  a factor of 3 - 6 smaller
than the typical "isoscalar" matrix elements in A=14-20 nuclei
(e.g. $\sim$ 0.3 eV in $^{19}$F \cite{AH} and $^{14}$N
\cite{HCBW93}). A analysis of the different
contributions to the PNC matrix elements indicates the
reasons for this behavior:

\noindent
i.) The one body transition densities (OBTD) for the isoscalar and
isovector $1d_{3/2}-2p_{3/2}$, $2s_{1/2}-2p_{1/2}$,
$1d_{5/2}-1f_{5/2}$ PNC transitions are a factor 5-10 smaller
than the dominant $1p_{1/2}-2s_{1/2}$ transition in the A=14-20
nuclei. This can be understood by the fact that for A=36 nuclei the
most probable subshell to be filled in the $1f2p$ major shell is
$1f_{7/2}$ which is 2 MeV lower than
$2p_{3/2}$. This substantially decreases the occupation
probability of the
$2p_{3/2}\ 2p_{1/2}\ $ and $1f_{5/2}$ states which are important for
PNC transitions. In the A=14-20 region the $2s_{1/2}$ is nearly
degenerate with the $1d_{5/2}$ state and has a higher occupation
probability.
In the
$^{36}$Cl case the dominant OBTD is for the transition
$1d_{3/2}-2p_{3/2}$.
We have estimated the $2\hbar \omega -1\hbar\omega$
and $2\hbar \omega -3\hbar\omega$ contribution to this dominant transition
within the
$(2s_{1/2},1d_{3/2},2p_{3/2},1f_{7/2})^{8}$ mode space.
The $2\hbar \omega -1\hbar\omega$
and
$2\hbar \omega
-3\hbar\omega$ contribution have a magnitude
of 5 - 7 \% of the $0\hbar \omega -1\hbar\omega$ and are
opposite in sign, so they are small and also
cancel each other. These calculations
suggest that the $0\hbar \omega -1\hbar\omega$ contribution is the
dominant term. This behaviour is in contrast with that in A=18-21 nuclei,
for which the higher $n\hbar \omega$ contributions seems to be important
\cite{HB94}. One can understand this by the relatively weaker coupling
between the sd and fp shells as compared with the coupling between the p
and sd shells.

\noindent
ii.) In the case of $^{36}$Cl, the isoscalar and isovector
contributions have opposite signs leading to a large suppression of the
total PNC matrix element (for the DDH weak coupling
constants). Considering the strong constraints on $h^{1}_{\pi}$
given by the $^{18}$F experiments, i.e. $h^{1}_{\pi} \leq (1/4)
(h^{1}_{\pi})_{DDH}$ \cite{PAGE87}, one obtains for
$(M_{PNC}^{^{36}Cl})_{DDH'}$ a value of -0.057 eV.

Taking into account the results of the above discussion,
we have used
$M^{^{36}Cl}_{PNC}$ = -0.057 eV and $M^{^{36}Ar}_{PNC}$ = 0.122 eV
to calculate the $(P_{\gamma})_{0}$.
We obtain $(P_{\gamma})_{0}^{^{36}Cl}$ = $1.1 \cdot 10^{-4}$ and
$(P_{\gamma})_{0}^{^{36}Ar}$ = $3.5 \cdot 10^{-4}$ (we used the
small mixing ratios, which agree with our calculations).
These values can be favorably compared with the experimental upper
limit, $P_{\gamma} \leq 3.9\cdot 10^{-4}$, for $^{18}$F \cite{PAGE87}.

One can try to avoid the low efficiency of the Compton polarimeters by
measuring the forward-backward asymmetry, Eqs. (\ref{eq:fbad}),
(\ref{eq:fbaa}). In this case one must find an efficient
polarization transfer mechanism which would permit one to obtain a PC
circular polarization $R_{\gamma}^{PC}$ larger than the polarimeter
efficiency ($\sim$ 1\% \cite{AH}).
For example, the $^{36}$Ar PMD can be
populated in the $^{39}K ( \vec{p}, \alpha ) ^{36}Ar$
reaction (in analogy with the $^{19}$F case \cite{SEATTLE}, \cite{ZUERICH}),
with low energy (E$_{p}$ $\geq$ 3.7 MeV) protons, while
the $^{36}$Cl PMD can be
populated in the $^{39}K ( \vec{n}, \alpha ) ^{36}Cl$
reaction with relatively slow (E$_{n}$ $\geq$ 0.6 MeV)
neutrons.

In conclusion, we have theoretically analyzed two new PMD cases in
mass A=36 nuclei. The parity nonconserving transition
for the PMD in $^{36}$Ar is isoscalar and the corresponding PNC observable
is sensitive to the dominant $h_{\rho}^{0}$ weak coupling constant, analogous
to the $0^{+}1$, $0^{-}1$ doublet in $^{14}$N \cite{HCBW93}.
The parity nonconserving transition
for the PMD in $^{36}$Cl is isoscalar+isovector and the corresponding
PNC observable is sensitive to the combination of $h_{\rho}^{0}$ and
$h_{\pi}^{1}$. From this point of view it is analogous to the $^{19}$F case.
However, the possible informations extracted from this case could be
complementary to the $^{19}$F result. Here, the isoscalar and isovector
contributions are out of phase, while for $^{19}$F they are in phase.
 The predicted circular polarizations of the order of magnitude
$10^{-4}$ are within the limits of accuracy of the existent experimental
setups.

\vspace{1.5cm}


{\large \bf Acknowledgments}
\vspace*{1.cm}

The author would like to thank Drs. B. Alex Brown  and O. Dumitrescu
for very  useful discussions during the preparation of the manuscript.
He want to acknowledge support from the
Alexander von Humboldt Foundation and NSF grant 94-03666.
He also thanks
Soros Foundation, Bucharest, Romania for a travel
grant.

\newpage

\begin{center}
{\bf Table captions}
\end{center}

{\bf Table 1} \ Input data, physical quantities and theoretical
PNC matrix elements necessary for calculating $\gamma$ - circular
polarizations and asymmetries for the two PMD - cases studied in
the present work.
The experimental data is taken from Ref. \cite{END} unless noted.\\

{\bf Table 2} \
Weak meson - nucleon coupling constants calculated
within different weak interaction models ( in units of 10$^{-7}$ ).
The abbreviations are: KM = Kaiser and Meissner \cite{KM}, DDH =
Desplanques, Donoghue and Holstein \cite{DDH}, AH = Adelberger and
Haxton \cite{AH} and DZ = Dubovik and Zenkin \cite{DZ}.
The $g_{meson}$ needed to obtain these results are
were taken from Ref.
\cite{AH}.\\

\vspace{1cm}

\begin{center}
{\bf Figure captions}
\end{center}

\vspace{1cm}

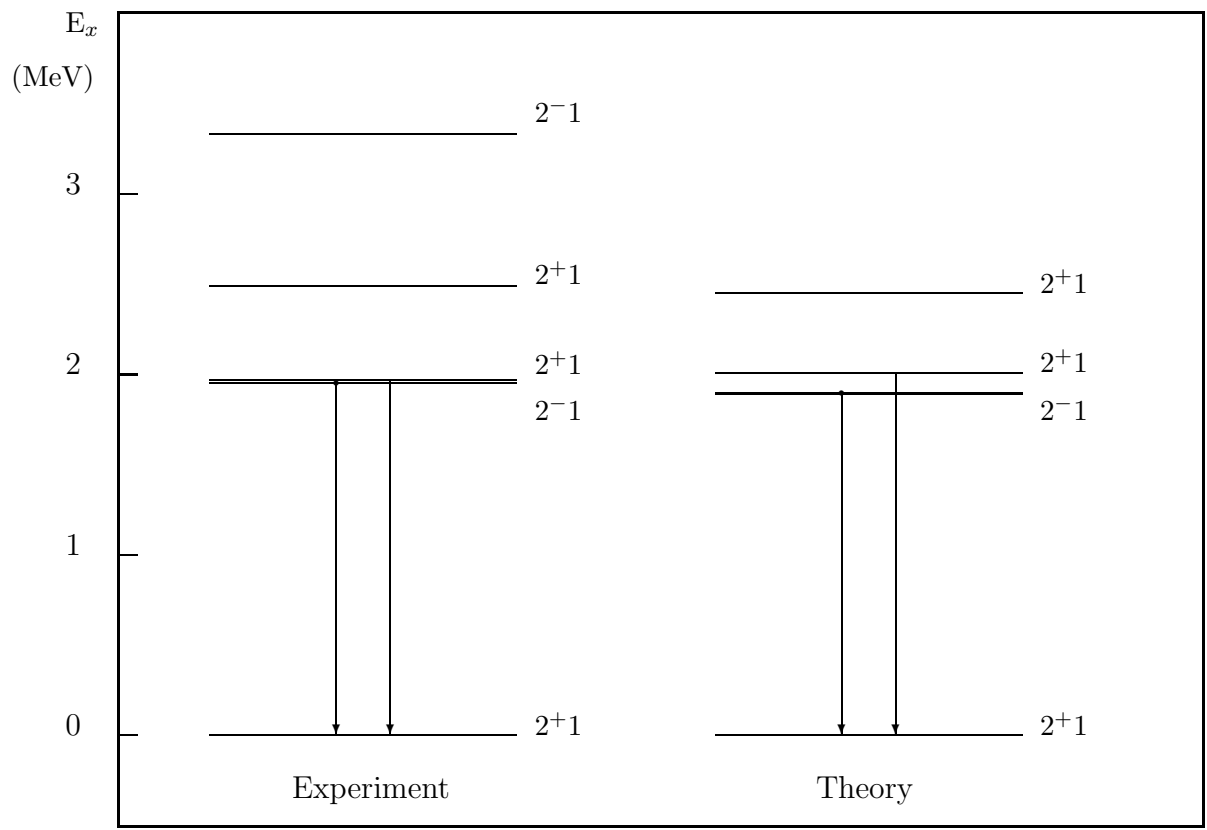
{\bf Figure 1} \ Experimentally and theoretically calculated
energies for the low 2$^{\pm}$1 levels in $^{36}$Cl. The first
excited $2^{+}1$ level has been artificially drawn 8 keV higher in
order that the PMD to be seen. \\

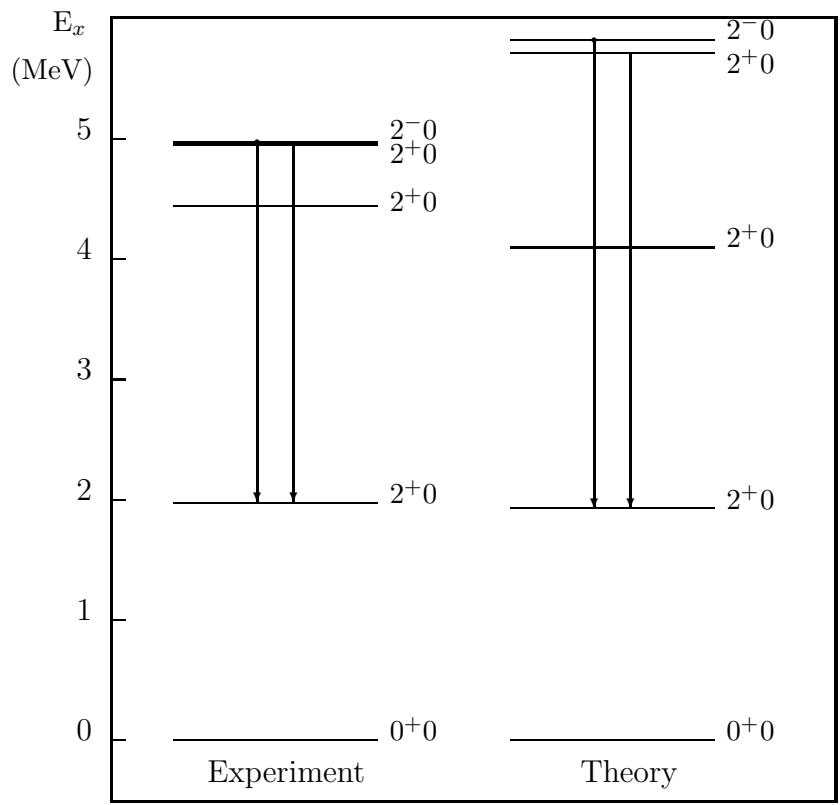
{\bf Figure 2} \ Same as Figure 1
for the low 2$^{\pm}$0 levels in $^{36}$Ar.\\

\newpage

\vspace{0.5cm}
\begin{center}
\begin{tabular}{|l|l|l|}  \hline
         &  $^{36}Cl$ &    $^{36}Ar$    \\
\hline
\hline
$I^{\pi}_{i} T_{i}, E_{i} (MeV)$ $\rightarrow$ &
$2^{+} 1, 1.959 MeV$ $\rightarrow$ &
$2^{+} 0, 4.951 MeV$ $\rightarrow$
\\
$I^{\pi}_{f} T_{f}, E_{f} (MeV) $ &
$2^{+} 1, g.s.$ &
$2^{+} 0, 1.97 MeV$
\\  \hline

life time ($\tau_{+}$)&
60 $\pm$ 15 fs &   $\leq$ 50 fs
\\  \hline

branching ratio ($b_{+}$) & 94.4 $\%$ & 15 $\%$
\\  \hline

mixing ratio ($\delta_{+}$)$_{exp}$ &
(-5.2 $\pm$ 1.6) or &
\\
 & (- 0.19 $\pm$ 0.06) \cite{KOPECKY} &
\\  \hline

mixing ratio ($\delta_{+}$)$_{theo}$ &
 -0.24 & 0.41
\\  \hline
B(M1)$_{exp}$
$\mu_{N}^{2}$
& 0.08 ($\delta_{+}$=-0.2);0.003 ($\delta_{+}$=-5.2) &
\\  \hline
B(M1)$_{theo}$
$\mu_{N}^{2}$
& 0.14 & 0.0009
\\  \hline
B(E2)$_{exp}$ e$^2$ fm$^4$  & 12 ($\delta_{+}$=-0.2); 298 ($\delta_{+}$=-5.2) &
\\  \hline
B(E2)$_{theo}$ e$^2$ fm$^4$ & 30 & 0.27\\
\hline
\hline

$I^{\pi}_{i} T_{i}, E_{i} (MeV)$ $\rightarrow$ &
$2^{-} 1, 1.951 MeV$ $\rightarrow$ &
$2^{-} 0, 4.974 MeV$ $\rightarrow$
\\
$I^{\pi}_{f} T_{f}, E_{f} (MeV) $ &
$2^{+} 1, g.s.$ &
$2^{+} 0, 1.97 MeV$
\\  \hline

life time ($\tau_{-}$) &  2.6 $\pm$ 0.3 ps & 14 $\pm$ 5 ps
\\  \hline

branching ratio ($b_{-}$) & 60 $\%$ & 4.0 $\pm$ 0.9 $\%$
\\  \hline

mixing ratio ($\delta_{-}$)$_{exp}$ &
(- 0.10 $\pm$ 0.10) \cite{KOPECKY} &
\\  \hline

mixing ratio ($\delta_{-}$)$_{theo}$ &
0.009 &
\\  \hline
B(E1)$_{exp}$ (e$^2$ fm$^2$)  & 1.9$\cdot$10$^{-5}$ & 0.7 $\cdot$ 10$^{-7}$
(if $\delta_{-}$ = 0.0)
\\  \hline
B(E1)$_{theo}$ (e$^2$ fm$^2$) &0.008 &
\\  \hline
B(M2)$_{exp}$
$\mu_{N}^{2}$ fm$^2$
& $\leq$ 25 &
\\  \hline
B(M2)$_{theo}$
$\mu_{N}^{2}$ fm$^2$
& 2.5 & 0.24
\\
\hline
\hline

$M^{DDH}_{PNC}$ (eV) & -0.019 &  0.122
\\  \hline
$M^{DDH}_{PNC}$ (eV), $h^{1}_{\pi}=\frac{1}{4}(h^{1}_{\pi})_{DDH}$
 & -0.057 &  0.122
\\  \hline
$M^{KM}_{PNC}$ (eV) & -0.023 &  0.067
\\  \hline
\end{tabular}

\vspace{.8cm}
Table 1.

\end{center}

\newpage

\vspace*{3.5cm}

\begin{center}
\vspace{0.5cm}

\begin{tabular}{|l|l|l|l|l|}  \hline
$h^{\Delta T}_{meson}$ &  KM   &    DDH  &   AH   &   DZ
\\  \hline
$h^{1}_{\pi}$          &  0.19 &   4.54  &   2.09 &  1.30
\\  \hline
$h^{0}_{\rho}$         & -3.70 & -11.40  &  -5.77 & -8.30
\\  \hline
$h^{1}_{\rho}$         & -0.10 &  -0.19  &  -0.22 &  0.39
\\  \hline
$h^{2}_{rho}$          & -3.30 &  -9.50  &  -7.06 & -6.70
\\  \hline
$h^{1}_{\rho^{'}}$     & -2.20 &   0.00  &   0.00 &  0.00
\\  \hline
$h^{0}_{\omega}$       & -1.40 &  -1.90  &  -4.97 & -3.90
\\  \hline
$h^{1}_{\omega}$       & -1.00 &  -1.10  &  -2.39 & -2.20
\\  \hline
\end{tabular}

\vspace{.8cm}

Table 2.

\end{center}

\newpage

\vspace*{3.5cm}

\begin{center}

\begin{minipage}[t]{16cm}
\unitlength2.4cm

\begin{picture}(6.5,5)(0,0)
\thicklines
\put(.5,.5){\framebox(6,4.5)}
\thinlines
\put(.2,4.9){\small E$_{x}$}
\put(-.1,4.6){\small (MeV)}
\put(1,1){\line(1,0){1.7}}
\put(3.8,1){\line(1,0){1.7}}
\put(1,2.951){\line(1,0){1.7}}
\put(1,2.968){\line(1,0){1.7}}
\put(1,3.492){\line(1,0){1.7}}
\put(1,4.332){\line(1,0){1.7}}
\put(3.8,3.008){\line(1,0){1.7}}
\put(3.8,3.451){\line(1,0){1.7}}
\put(3.8,2.894){\line(1,0){1.7}}
\put(1.4,.65){ Experiment}
\put(4.3,.65){ Theory}
\put(1.7,2.951){\vector(0,-1){1.951}}
\put(2.,2.968){\vector(0,-1){1.968}}
\put(4.5,2.894){\vector(0,-1){1.894}}
\put(4.8,3.008){\vector(0,-1){2.008}}
\put(1.7,2.951){\circle*{.03}}
\put(4.5,2.894){\circle*{.03}}
\put(2.8,1){\small 2$^{+}$1}
\put(2.8,2.75){\small 2$^{-}$1}
\put(2.8,3.0){\small 2$^{+}$1}
\put(2.8,4.4){\small 2$^{-}$1}
\put(2.8,3.5){\small 2$^{+}$1}
\put(5.6,1){\small 2$^{+}$1}
\put(5.6,2.75){\small 2$^{-}$1}
\put(5.6,3.01){\small 2$^{+}$1}
\put(5.6,3.451){\small 2$^{+}$1}

\put(.5,1){\line(1,0){.1}}
\put(.5,2){\line(1,0){.1}}
\put(.5,3){\line(1,0){.1}}
\put(.5,4){\line(1,0){.1}}
\put(.2,1){0}
\put(.2,2){1}
\put(.2,3){2}
\put(.2,4){3}

\end{picture}

\begin{center}
Figure 1.
\end{center}

\end{minipage}

\end{center}

\newpage

\vspace*{3.5cm}

\begin{center}

\begin{minipage}[t]{12cm}
\unitlength1.6cm

\begin{picture}(6.5,7)(0,0)
\thicklines
\put(.5,.5){\framebox(6,6.5)}
\thinlines
\put(.0,6.9){\small E$_{x}$}
\put(-.35,6.5){\small (MeV)}
\put(1,1){\line(1,0){1.7}}
\put(3.8,1){\line(1,0){1.7}}

\put(1,2.970){\line(1,0){1.7}}
\put(1,5.444){\line(1,0){1.7}}
\put(1,5.951){\line(1,0){1.7}}
\put(1,5.974){\line(1,0){1.7}}

\put(3.8,2.927){\line(1,0){1.7}}
\put(3.8,5.097){\line(1,0){1.7}}
\put(3.8,6.713){\line(1,0){1.7}}
\put(3.8,6.823){\line(1,0){1.7}}

\put(1.2,.65){ Experiment}
\put(4.3,.65){ Theory}

\put(1.7,5.974){\vector(0,-1){3.004}}
\put(2.,5.951){\vector(0,-1){2.981}}
\put(4.5,6.823){\vector(0,-1){3.896}}
\put(4.8,6.713){\vector(0,-1){3.786}}
\put(1.7,5.974){\circle*{.05}}
\put(4.5,6.823){\circle*{.05}}
\put(2.8,1){\small 0$^{+}$0}
\put(2.8,2.97){\small 2$^{+}$0}
\put(2.8,5.4){\small 2$^{+}$0}
\put(2.8,6.){\small 2$^{-}$0}
\put(2.8,5.8){\small 2$^{+}$0}
\put(5.6,1){\small 0$^{+}$0}
\put(5.6,2.927){\small 2$^{+}$0}
\put(5.6,5.097){\small 2$^{+}$0}
\put(5.6,6.55){\small 2$^{+}$0}
\put(5.6,6.83){\small 2$^{-}$0}

\put(.5,1){\line(1,0){.1}}
\put(.5,2){\line(1,0){.1}}
\put(.5,3){\line(1,0){.1}}
\put(.5,4){\line(1,0){.1}}
\put(.5,5){\line(1,0){.1}}
\put(.5,6){\line(1,0){.1}}
\put(.2,1){0}
\put(.2,2){1}
\put(.2,3){2}
\put(.2,4){3}
\put(.2,5){4}
\put(.2,6){5}

\end{picture}

\begin{center}
Figure 2.
\end{center}

\end{minipage}

\end{center}

\end{document}